\documentclass[journal,transmag]{IEEEtran}

\usepackage{cite}

\usepackage{graphicx}
\ifCLASSINFOpdf
\else
\fi

\usepackage{amsmath,amssymb}
\usepackage{algorithmic}

\usepackage{array}

\ifCLASSOPTIONcompsoc
\usepackage[caption=false,font=normalsize,labelfont=sf,textfont=sf]{subfig}
\else
  \usepackage[caption=false,font=footnotesize]{subfig}
\fi

\usepackage{url}

\begin{document}
\IEEEoverridecommandlockouts
\IEEEpubid{\begin{minipage}{\textwidth}\vspace{20pt}
		\fbox{\parbox{\dimexpr\textwidth-2\fboxsep-2\fboxrule\relax}{\small
				© 2026 IEEE. Personal use of this material is permitted. Permission from IEEE must be obtained for all other uses, in any current or future media, including reprinting/republishing this material for advertising or promotional purposes, creating new collective works, for resale or redistribution to servers or lists, or reuse of any copyrighted component of this work in other works.}}\vspace{-32pt}
\end{minipage}}
\title{Simplification of the Isotropic Generalized Stop-Type Prandtl-Ishlinskii Vector Hysteresis Operator Using \\Analytical Return-Point Mapping}

\author{\IEEEauthorblockN{Arvinth Shankar\IEEEauthorrefmark{1,2},
		Klaus Kuhnen\IEEEauthorrefmark{1}, Iryna Kulchytska-Ruchka\IEEEauthorrefmark{1}, Sebastian Schöps\IEEEauthorrefmark{2}}
\IEEEauthorblockA{\IEEEauthorrefmark{1}Robert Bosch GmbH,
	Bosch Research and Advance Engineering, Germany}
\IEEEauthorblockA{\IEEEauthorrefmark{2}Computational Electromagnetics Group, Technische Universität Darmstadt, Germany}}

\IEEEtitleabstractindextext{%
\begin{abstract}
While the thermodynamically formulated generalized Prandtl--Ishlinskii stop-type operator effectively captures hysteresis nonlinearities, it requires a local iterative procedure to update each hysteron, resulting in considerable computational effort. In this work, we propose a simplified thermodynamic formulation of the generalized Prandtl--Ishlinskii stop operator. The nonlinear mapping on the stop operator is replaced by an identity, such that the hysteresis operators are directly weighted through their outputs, while the nonlinear anhysteretic response, represented by ramp dead-zone basis functions, is fully preserved. For isotropic cases, this simplification enables a closed-form solution for the local plastic correction, eliminating per-hysteron iterative Newton updates. The resulting constitutive mapping is integrated into a finite element solver, and numerical results show a significant reduction in computation time with accuracy comparable to the generalized model.
\end{abstract}

}

\maketitle

\IEEEdisplaynontitleabstractindextext

\IEEEpeerreviewmaketitle

\section{Introduction}
\IEEEPARstart{R}{eal} ferromagnetic materials exhibit complex hysteresis effects, requiring memory-dependent constitutive mapping between the magnetic field~$\mathbf{H}$ and the flux density~$\mathbf{B}$ for solving low-frequency Maxwell's equations. Operator-based phenomenological approaches \cite{Bobbio:playStopHyst} are widely used, expressed as a vector-valued, rate-independent operator $\mathbf{H} = \textbf{\textit{V}}[\mathbf{B}]$. The generalized Prandtl--Ishlinskii model~\cite{Saaideh:genPrandtl} effectively characterizes hysteresis nonlinearities and its stop-type formulation defines the mapping $\textbf{\textit{V}}$ through a weighted superposition of generalized stop operators. Its rheological representation~\cite{Krejci:rheologyThermo, Visintin:rheology} combines a nonlinear elastic spring in parallel with multiple generalized vector-stop hysterons, and a thermodynamic interpretation of such nonlinear vector-stop elements for anisotropic materials is discussed in~\cite{Xiao:anisotropic}. Thermodynamically formulated models provide physically meaningful constructions and allows decomposition of dissipative and reactive power~\cite{Krejci:rheologyThermo}.
 
In the thermodynamically formulated generalized Prandtl--Ishlinskii stop-type model~\cite{Xiao:anisotropic}, each hysteron is split into an elastic (reversible) part and a plastic (irreversible) part, where the plastic part constitutes the internal memory state. The hysteron output is a nonlinear function of the elastic part, yielding an implicit update equation for each hysteron and can only be resolved iteratively through the Newton method. This local, nonlinear, iterative update incurs considerable computational effort, particularly in finite element (FE) simulations.

Although $\textbf{\textit{V}}$ is local and parallelizable across integration points, the per-point hysteretic updates remain expensive.
In this work, we propose a simplification of the generalized Prandtl--Ishlinskii stop operator within the thermodynamic framework, while retaining the essential hysteresis properties. This simplification linearly weights the stop-operator outputs, and for isotropic cases, this yields an analytical return-point mapping for the local hysteron updates, eliminating the need for iterative Newton procedures. Integrated into an FE framework, the simplified constitutive mapping enables faster evaluation of~$\textbf{\textit{V}}[\mathbf{B}]$ at each integration point, reducing the overall computational cost.

The paper is organized as follows. In Sec.~\ref{section2a}, we introduce the operator simplification and its rheological interpretation. The anhysteretic and hysteretic formulations are presented in~Sec.~\ref{section2b} and Sec.~\ref{section2c}, respectively. Thermodynamic 
consistency of the simplification is verified in Sec.~\ref{section2d}, and parameter identification 
is briefly discussed in Sec.~\ref{section2e}. In Sec.~\ref{section3}, we present a 2D FE simulation comparing computational performance 
and losses, and finally conclude in Sec.~\ref{section4}.

\section{Simplified model and its formulation}
\label{section2}
\subsection{Operator simplification}
\label{section2a}
\begin{figure}[!t]
	\centering
	\subfloat[Generalized model]{
		\includegraphics[width=0.46\columnwidth]{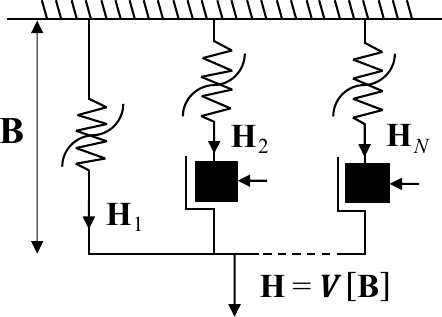}
		\label{fig_generalized_model}
	}\hfil
	\subfloat[Simplified model]{
		\includegraphics[width=0.46\columnwidth]{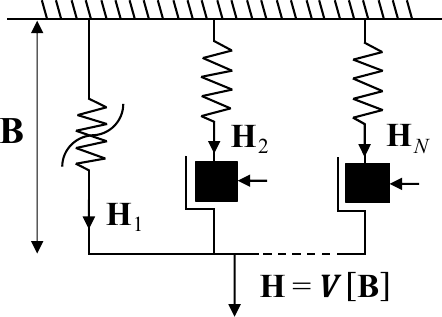}
		\label{fig_simplified_model}
	}
	\caption{Prandtl--Ishlinskii stop-type rheological network.}
	\label{fig_PrandtlIshlinskiiStopTypeModels}
\end{figure}

The generalized Prandtl--Ishlinskii stop-type operator defines the mapping between the output field intensity $\mathbf{H}$ and input flux density $\mathbf{B}$ in a discrete formulation~\cite{Saaideh:genPrandtl,Xiao:anisotropic}  as
\begin{equation}
	\mathbf{H} = \textbf{\textit{V}}[\mathbf{B}] =
	 w_1\,g(\mathbf{B}) +
	\sum_{i=2}^{N} w_i\,g(\textbf{\textit{S}}_i[\mathbf{B}]),
	\label{eq:GeneralizedPrandtl}
\end{equation}
where $g$ is a nonlinear memoryless mapping, $\textbf{\textit{S}}_i$ are the memory-dependent operators of $i$-th stop hysteron, $w_1$ and $w_i$ are the weight coefficients, and $N$ denotes the total number of rheological elements (Fig.~\ref{fig_generalized_model}). The first term represents the anhysteretic response, while the summation accounts for the hysteretic contribution through a weighted superposition of $g$ applied to the operators~$\textbf{\textit{S}}_i$. 
Within the thermodynamic framework, the memoryless mapping is identified as~$g(\cdot)=\nabla f(\cdot)$, where~$f$ is the Helmholtz free energy potential.
Hence, each hysteron contribution takes the nonlinear form $\nabla f$ applied to the stop operators $\textbf{\textit{S}}_i$. We introduce a simplification of~\eqref{eq:GeneralizedPrandtl}, retaining the anhysteretic term $w_1\nabla f(\mathbf{B})$ while simplifying the hysteretic contribution by treating~$\nabla f$ as an identity mapping. The stop hysterons are thus weighted directly through their outputs $\textbf{\textit{S}}_i$ as follows
\begin{equation}
	\mathbf{H} = \textbf{\textit{V}}[\mathbf{B}] =
	w_1 \nabla f(\mathbf{B}) +
	\sum_{i=2}^{N}  w^\mathrm{new}_{i}\,\textbf{\textit{S}}_i[\mathbf{B}].
	\label{eq:SimplifiedPrandtlThermodynamical}
\end{equation}
Here,  $w^\mathrm{new}_{i}$ are refitted hysteron weights. In the rheological network, this is equivalent to replacing the nonlinear spring in each hysteron with a linear spring of unit stiffness ($1\,$N/m) as shown in Fig.~\ref{fig_simplified_model}.

\subsection{Anhysteretic characteristic}
\label{section2b}
\begin{figure}[!b]
	\centering
	\subfloat[Zero dead-zone threshold]{\includegraphics[width=0.44\columnwidth]{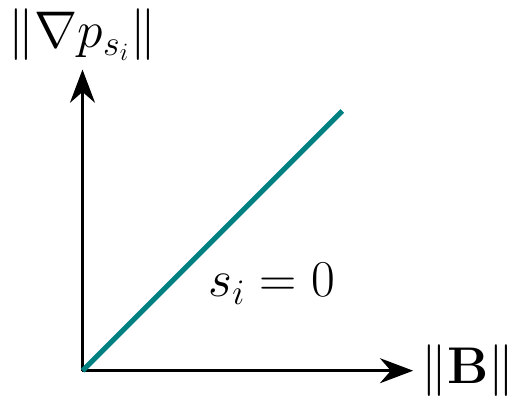}
		\label{fig:zeroThreshold}}
	\hfil
	\subfloat[Nonzero dead-zone threshold]{\includegraphics[width=0.44\columnwidth]{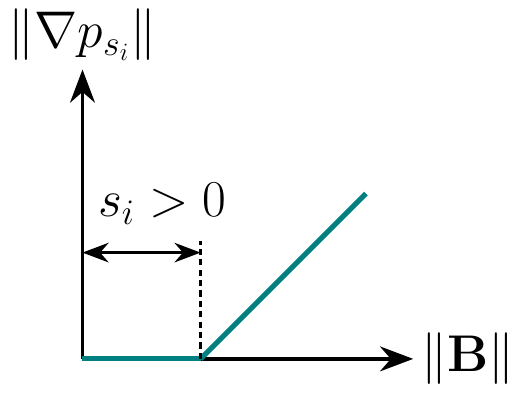}
		\label{fig:nonzeroThreshold}}
	\caption{Basis potential gradients as a 
		function of magnetic flux density.}
	\label{fig:dead_zone_thresholds}
\end{figure}

\begin{figure}[!ht]
	\centering
	\subfloat[Dirac delta hat function]{\includegraphics[width=0.36\columnwidth]{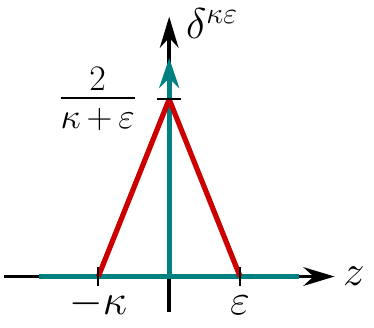}
		\label{fig:DiracDelta}}
	\hfil
	\subfloat[Regularized basis potential gradient]{\includegraphics[width=0.38\columnwidth]{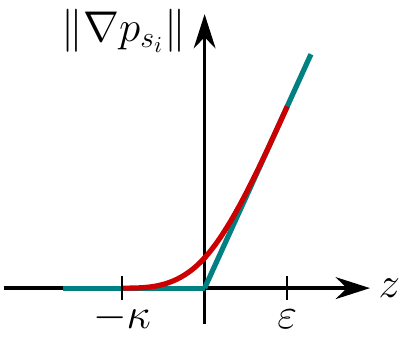}
		\label{fig:potGradReg}}
	\caption{Regularized basis potential gradient using Dirac delta hat function.}
	\label{fig:diracAndReg}
\end{figure}

The spring-only branch is modeled by the isotropic free
energy potential~$f$. It is expressed as a linear combination of $m$ piecewise basis potentials~$p_{s_i}\in\mathbb{P}_2$ that depend on flux density~$\mathbf{B}$, with weights $v_i$ as follows
\begin{equation}
	f(\mathbf{B}) = \sum_{i=1}^{m} v_i\,p_{s_i}. \label{eq:freeEnergy}
\end{equation}
Here, $s_i$ denote the dead-zone thresholds~\cite{Kuhnen:modifiedPrandtl}.
The potential gradient $\nabla f$ and the Hessian $\nabla^2 f$ are further obtained as
\begin{equation}
	\nabla f(\mathbf{B}) = \sum_{i=1}^{m} v_i \nabla p_{s_i}\,, \qquad
	\nabla^2 f(\mathbf{B}) = \sum_{i=1}^{m} v_i \nabla^2 p_{s_i}.
	\label{eq:potentialGradHessian}
\end{equation}
As shown in Fig.~\ref{fig:dead_zone_thresholds}, a zero dead-zone threshold $s_i$ yields a linear basis potential gradient
while a nonzero threshold introduces a piecewise linear characteristic.
The anhysteretic response $\nabla f$ is thus constructed as a weighted superposition of such piecewise linear basis functions.
To ensure smoothness of the gradient and Hessian for numerical robustness, a regularization is applied using the Dirac delta hat function $\delta^{\kappa\varepsilon}$ (Fig.~\ref{fig:DiracDelta}), where $\kappa$ and $\varepsilon$ are the regularization parameters. The coordinate shift $z = \|\mathbf{B}\|-s_i$ is introduced  to relocate the $i$-th threshold to the origin.
Integrating $\delta^{\kappa\varepsilon}$ twice yields $C^2$ continuous basis potential gradients, as illustrated in Fig.~\ref{fig:potGradReg}.

\subsection{Hysteretic characteristic}
\label{section2c}
\begin{figure}[!t]
	\centering
	\subfloat[Generalized hysteron]{
		\includegraphics[width=0.44\columnwidth]{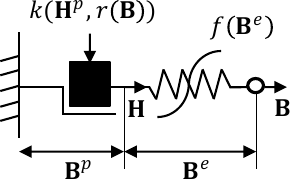}
		\label{fig_generalized_stop}
	}\hfil
	\subfloat[Simplified hysteron]{
		\includegraphics[width=0.44\columnwidth]{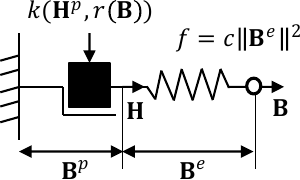}
		\label{fig_simplified_stop}
	}
	\caption{Vector stop model.}
	\label{fig_stop}
\end{figure}

\begin{figure}[!t]
	\centering
	\includegraphics[width=1.7in]{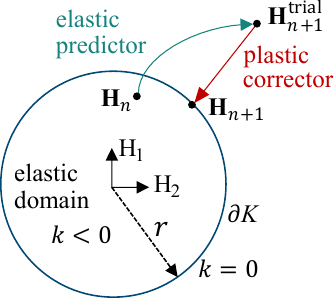}
	\caption{Elastic predictor--plastic corrector in 2D admissible domain.}
	\label{fig_admissibleDomain}
\end{figure}

The hysteretic response is produced by the stop hysteron (Fig.~\ref{fig_stop}). For this section, a representative hysteron is considered and the hysteron index $i$ in \eqref{eq:GeneralizedPrandtl}, \eqref{eq:SimplifiedPrandtlThermodynamical} is omitted for readability.
In the vector stop model, the flux density decomposes as $\mathbf{B} = \mathbf{B}^e+\,\mathbf{B}^p$, where $\mathbf{B}^e$, $\mathbf{B}^p$ are the elastic and plastic parts, respectively, while $\mathbf{H} = \mathbf{H}^e = \mathbf{H}^p$. The plastic element in both models (Fig.~\ref{fig_stop}) is described by a strongly convex quadratic dissipation
potential $k$, dependent on the plastic field~$\mathbf{H}^p$ and scalar threshold $r(\mathbf{B})$ as
\begin{equation}
		k(\mathbf{H}^p,r(\mathbf{B})) = \frac{1}{2} (\mathbf{H}^p)^\top \mathbf{H}^p - \frac{1}{2} r(\mathbf{B})^2 \leq 0.
	\label{eq:diss_pot}
\end{equation}
The threshold $r$ depends on the flux density to deactivate hysterons under large $\mathbf{B}$, where the response is nearly reversible.
The constraint $k\leq0$ defines a strongly convex admissible set 
\begin{equation}
	K(r(\mathbf{B})) = \left\{\mathbf{H}^p\,|\, k(\mathbf{H}^p,r(\mathbf{B})) \leq 0\right\},
	\label{eq:admissibleset}
\end{equation}
where $k < 0$ inside the elastic domain and $k=0$ on the boundary~$\partial K$ (Fig.~\ref{fig_admissibleDomain}). 
For isotropic cases, $\partial K$ forms a circle in 2D or a sphere in 3D with radius determined by threshold~$r$.
The maximum dissipation principle from plasticity theory~\cite{Hackl:MaxDissipationPrinciple} determines 
a unique plastic flow direction by finding the optimal hysteron field response $\mathbf{H}^*$ that maximizes the dissipative   
power~$\mathbf{H}^\top\dot{\mathbf{B}}^p$ subject to $k \leq 0$, i.e.
\begin{equation}
	\mathbf{H}^* = \arg\max_{\mathbf{H} \in K}
	\left(\mathbf{H}^\top\dot{\mathbf{B}}^p\right).
	\label{eq:MaximumDissipationPrinciple}
\end{equation}
Solving~\eqref{eq:MaximumDissipationPrinciple} yields the evolution equation
\begin{equation}
	\dot{\mathbf{B}}^p = \lambda\nabla k,
	\label{eq:evolutionEquation}
\end{equation}
with Lagrange parameter $\lambda$ and Karush-Kuhn-Tucker (KKT) conditions: $k \leq 0$, $\lambda \geq 0$ and $\lambda k = 0$. For the generalized hysteron, the nonlinear elastic spring gives $\mathbf{H} = \nabla f(\mathbf{B}^e)$ and~\eqref{eq:evolutionEquation} reduces to $\dot{\mathbf{B}}^p = \lambda\mathbf{H} = \lambda\nabla f(\mathbf{B} - \mathbf{B}^p)$.
Applying backward-Euler time discretization yields
\begin{equation}
	\mathbf{B}_{n+1}^p - \mathbf{B}_{n}^p = \eta_{n+1} \nabla f(\mathbf{B}_{n+1} - \mathbf{B}_{n+1}^p),
	\label{eq:GeneralizedDiscreteEvolutionEquation}
\end{equation}
where $\eta$ is the discrete-time Lagrange parameter and $n+1$ denotes the current time step.

For the simplified hysteron, a quadratic free energy potential with unit stiffness ($c=1$ in Fig.~\ref{fig_simplified_stop}) gives the hysteron response $\mathbf{H} = \mathbf{B}^e$. As a result,~\eqref{eq:evolutionEquation} reduces to $\dot{\mathbf{B}}^p = \lambda \mathbf{B}^e$.
Further time discretization yields the update equation
\begin{equation}
	\mathbf{B}_{n+1}^p - \mathbf{B}_{n}^p = \eta_{n+1} (\mathbf{B}_{n+1} - \mathbf{B}_{n+1}^p).
	\label{eq:SimplifiedDiscreteEvolutionEquation}
\end{equation}
Both the updates~\eqref{eq:GeneralizedDiscreteEvolutionEquation} and~\eqref{eq:SimplifiedDiscreteEvolutionEquation} are solved using an elastic predictor-plastic corrector return point mapping scheme~\cite[pp. 140--143]{Simo:compInelastic}. First, a trial state is computed based on previous plastic state $\mathbf{B}^p_n$, assuming no plastic evolution:
\begin{subequations}
	\begin{align}
		\mathbf{H}^\mathrm{trial}_{n+1} &= \nabla f(\mathbf{B}_{n+1}-\mathbf{B}^p_n), \\
		\mathbf{H}^\mathrm{trial}_{n+1} &= \mathbf{B}_{n+1}-\mathbf{B}^p_n, \label{eq:HtrialForSimplified}
	\end{align}
\end{subequations}
for the generalized and simplified models, respectively.
If $\|\mathbf{H}^\mathrm{trial}_{n+1}\| >  r(\mathbf{B}_{n+1})$, a plastic correction projects the trial state back onto~$\partial K$~(Fig.~\ref{fig_admissibleDomain}). In the generalized model, this requires iterative Newton solves due to the nonlinear free energy potential. In the simplified model, the linear elastic relation yields the fully analytical return-point mapping
\begin{equation}
	\mathbf{H}_{n+1} = r(\mathbf{B}_{n+1}) \frac{\mathbf{H}^\mathrm{trial}_{n+1}}{\|\mathbf{H}^\mathrm{trial}_{n+1}\|}\,,\quad
	\eta_{n+1} = \frac{\|\mathbf{H}^\mathrm{trial}_{n+1}\|}{r(\mathbf{B}_{n+1})} -1\,,
	\label{eq:returnPointMappingSimplified}
\end{equation}
enabling direct analytical hysteron updates without iteration. 
\subsection{Thermodynamic consistency}
\label{section2d}
The energy balance for the $i$-th hysteron decomposes the input power into stored and dissipated contributions as
\begin{equation}
	\mathbf{H}_i^\top\dot{\mathbf{B}}_i = \mathbf{H}_i^\top\dot{\mathbf{B}}_i^e + \mathbf{H}_i^\top\dot{\mathbf{B}}_i^p,
\end{equation}
where $\dot{\mathbf{B}}_i^e$ and $\dot{\mathbf{B}}_i^p$ are the elastic and plastic flux rates, respectively.
The hysteresis loss density $p[\mathbf{B}]$ is obtained as a weighted sum of dissipative contributions from all hysterons in Fig.~\ref{fig_simplified_model} as,
\begin{equation}
	p[\mathbf{B}] = \mathbf{H}^\top\dot{\mathbf{B}}^p = \sum_{i=2}^{N} w^\mathrm{new}_{i}\,\mathbf{H}_i^\top\dot{\mathbf{B}}_i^p.
	\label{eq:dissipativeTerm}
\end{equation}
Substituting $\mathbf{H}_i = \mathbf{B}^e_i$ from the simplified model and the evolution equation $\dot{\mathbf{B}}_i^p = \lambda \mathbf{B}_i^e$ into \eqref{eq:dissipativeTerm} yields
\begin{equation}
	\mathbf{H}^\top\dot{\mathbf{B}}^p = \sum_{i=2}^{N} w^\mathrm{new}_{i}\,{\mathbf{B}_i^e}^\top \lambda \mathbf{B}_i^e,
	\label{eq:finalDissipativeExpression}
\end{equation}
which involves two key quantities: $\lambda \geq 0$ from KKT conditions and ${\mathbf{B}_i^e}^\top\mathbf{B}_i^e \geq 0$. The non-negativity of weights $w^\mathrm{new}_{i}$ is enforced during parameter identification, ensuring overall non-negative dissipation and thus thermodynamic consistency of the simplified formulation.
\subsection{Parameter identification}
\label{section2e}
The first stage identifies the anhysteretic parameters through a nested nonlinear program for the dead-zone thresholds~$s_i$ and an inner quadratic program for the coefficients~$v_i$, both defined in~\eqref{eq:freeEnergy}.
Constraints on~$v_i$ ensuring strong convexity of potential~$f$ and a monotonous gradient are given in~\cite{Kuhnen:modifiedPrandtl}. 
Deep saturation follows by extrapolating the potential gradient to the reluctivity of free space. In the second stage, the hysteron weights $w^\mathrm{new}_{i}$ are obtained by minimizing the error between the overall model response and the measured field $\mathbf{H}_\mathrm{meas}$,
\begin{equation}
	\mathbf{e} = \sum_{i=1}^{N}w^\mathrm{new}_{i}\,\mathbf{H}_i - \mathbf{H}_\mathrm{meas},
	\label{eq:errorHysteronParameters}
\end{equation}
where $\mathbf{H}_i$ is the field response from $i$-th branch in Fig.~\ref{fig_simplified_model}. The optimal $N$~is identified where the error~\eqref{eq:errorHysteronParameters} saturates, yielding e.g. $N = 9$ for M330-35A in simplified model (Sec.~\ref{section3b}).
\section{Numerical results}
\label{section3}
\subsection{Comparison of $B$--$H$ curves and hysteresis loss density}
\label{section3a}
\begin{figure}[!t]
	\centering
	\includegraphics[width=2.55in]{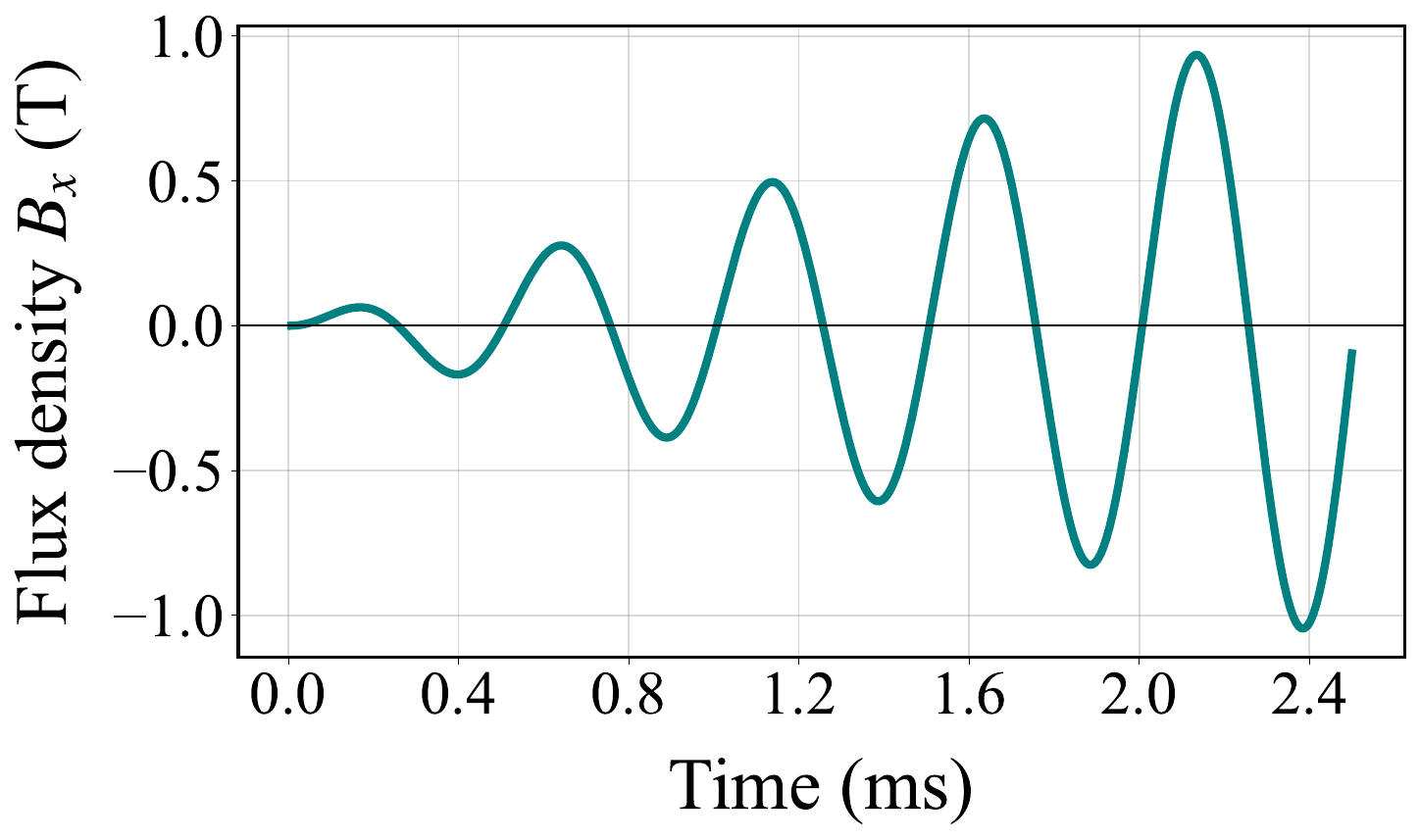}
	\caption{Input flux density $B_x$ as a function of time.}
	\label{fig_input_signal}
\end{figure}
\begin{figure}[!t]
	\centering
	\includegraphics[width=2.7in]{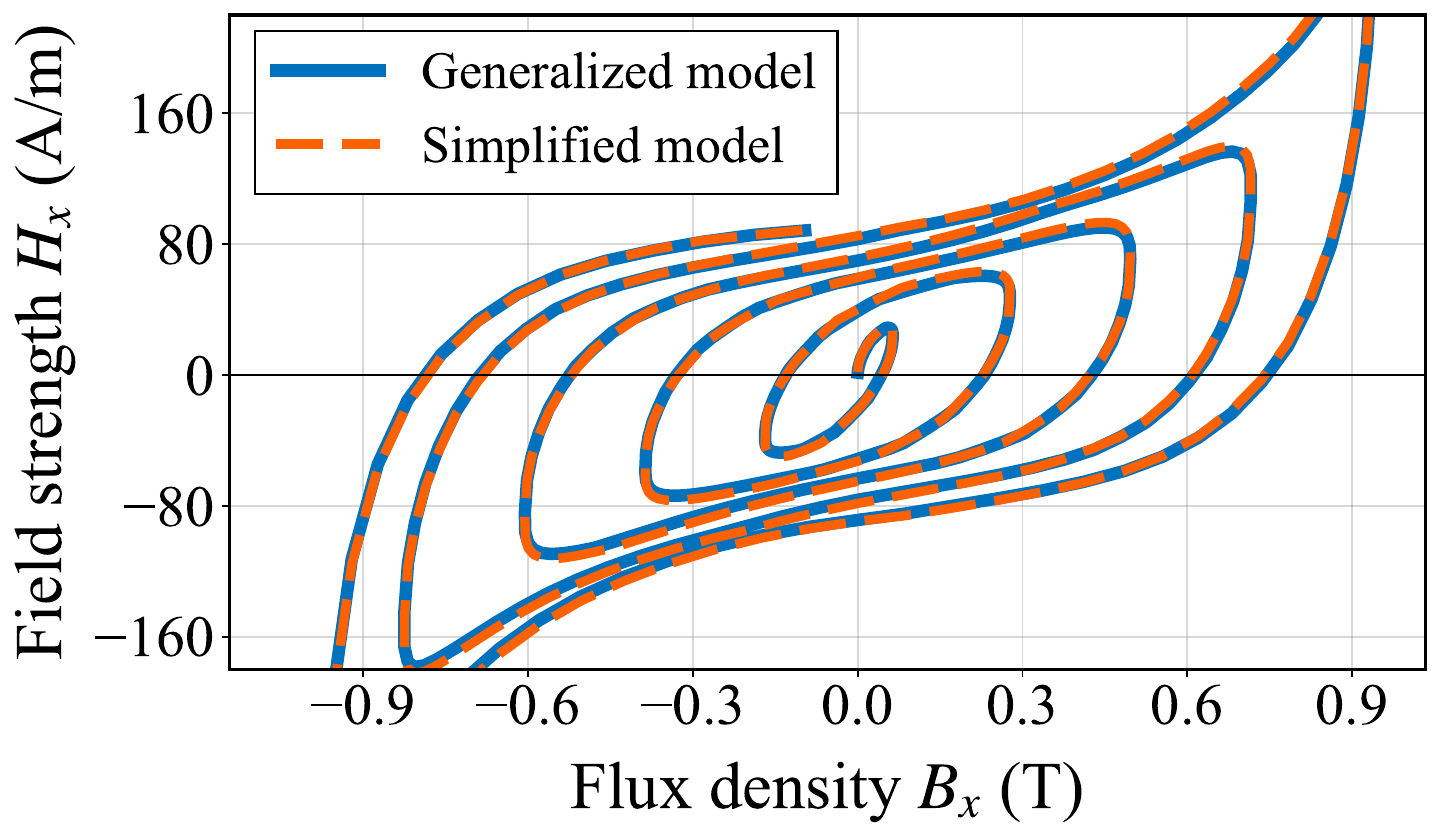}
	\caption{Comparison of $B_x$-$H_x$ responses.}
	\label{fig_bhCurve}
\end{figure}
\begin{figure}[!t]
	\centering
	\includegraphics[width=2.52in]{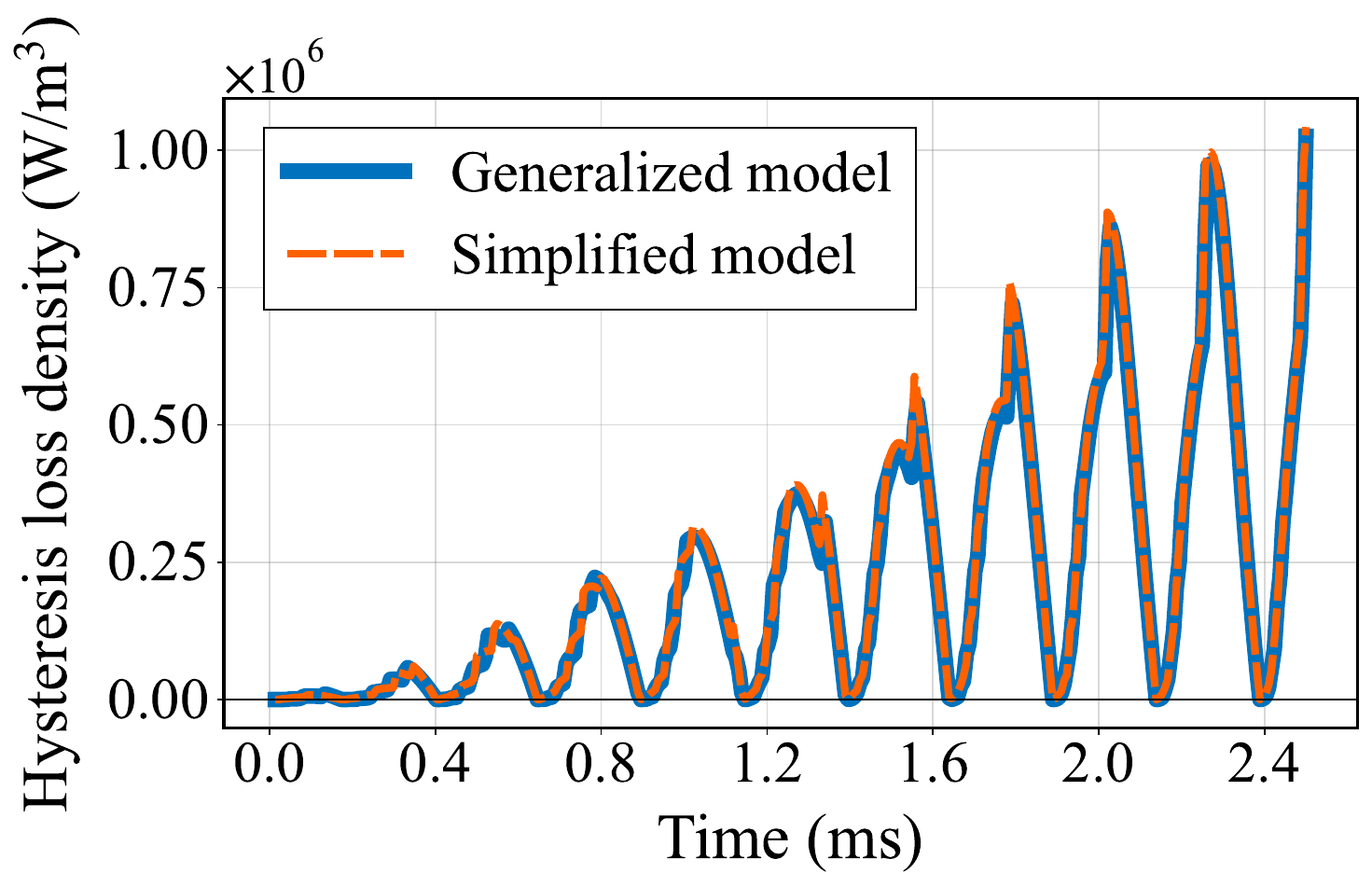}
	\caption{Comparison of hysteresis loss density.}
	\label{fig_hyst_loss}
\end{figure}

The material response is first assessed at a single point, driven by a 2D oscillating flux of increasing amplitude (Fig.~\ref{fig_input_signal}) in~$B_x$ and $B_y$
without field coupling. Both models produce nearly identical $B_x$--$H_x$ responses, capturing similar minor loop behavior across the excitation range (Fig.~\ref{fig_bhCurve}).
The hysteresis loss density comparison (Fig.~\ref{fig_hyst_loss}) yields an overall relative difference of approximately $4.71\%$, confirming comparable accuracy of the simplified model.

\subsection{FE Simulation of 2D electric machine}
\label{section3b}
\begin{figure}[!t]
	\centering
	\includegraphics[width=1.45in]{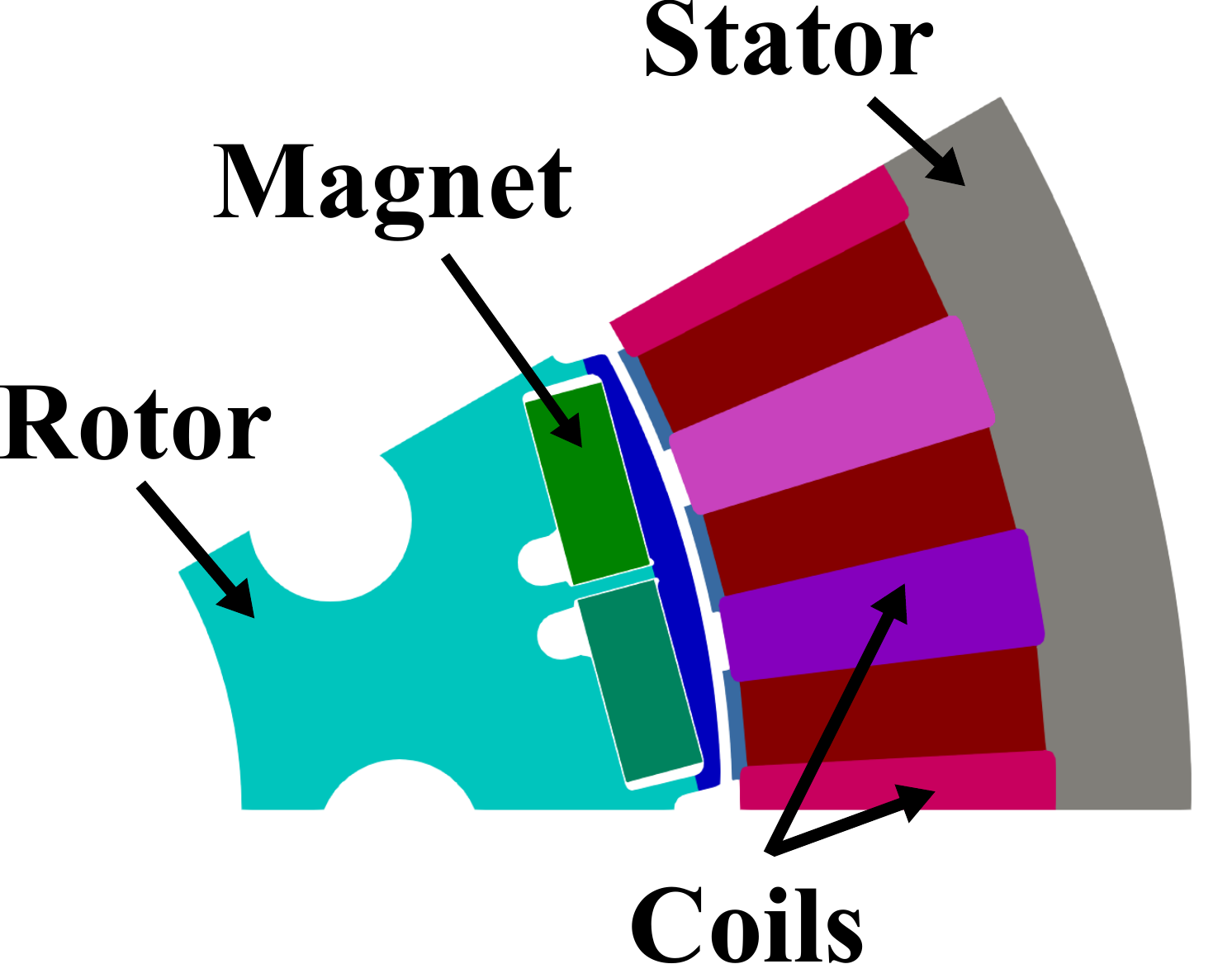}
	\caption{2D PMSM model.}
	\label{fig_2d_pmsm}
\end{figure}

\begin{figure}[!b]
	\centering
	\includegraphics[width=2.4in]{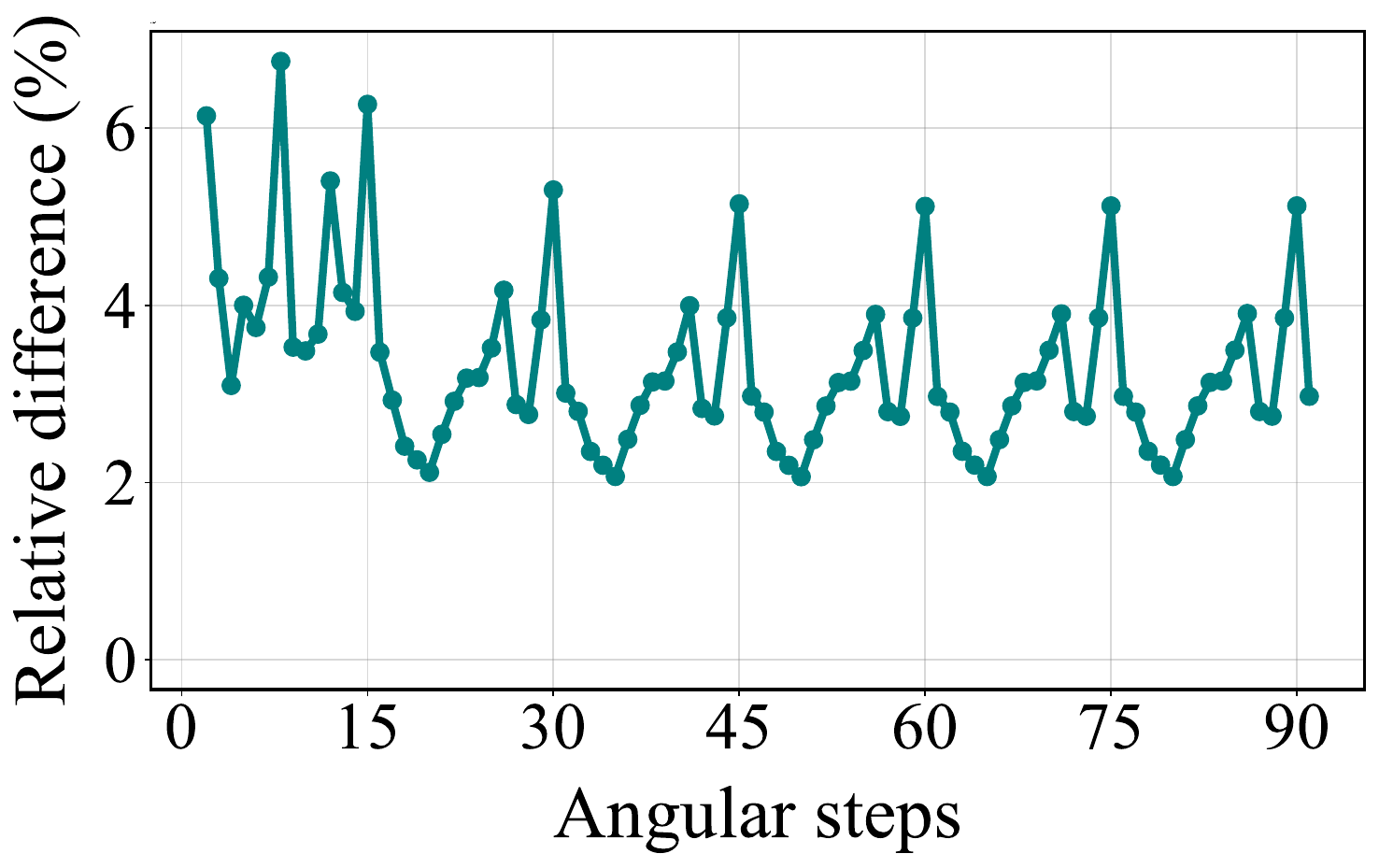}
	\caption{Relative difference ($\%$) for the instantaneous hysteresis loss (W).}
	\label{fig_hyst_loss_first_case}
\end{figure}

\begin{figure}[!t]
	\centering
	\includegraphics[width=2.5in]{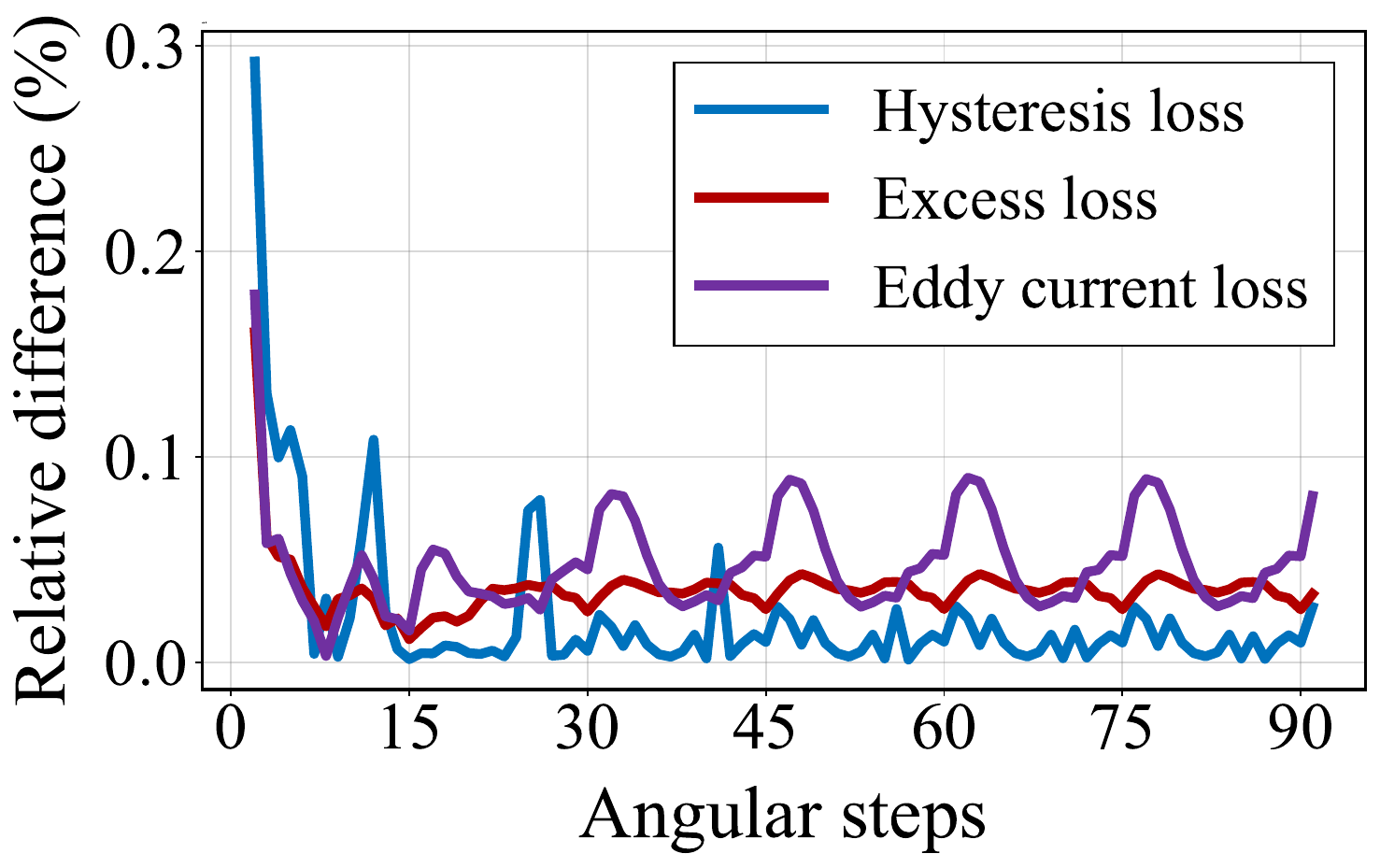}
	\caption{Relative difference ($\%$) for the instantaneous hysteresis, excess and eddy-current loss with respect to the generalized model using offline approach.}
	\label{fig_hyst_excess_and_eddy}
\end{figure}

Both models are then applied to a 2D Permanent Magnet Synchronous Machine (PMSM) FE model (Fig.~\ref{fig_2d_pmsm}), discretized with 10,000 elements. The stator and rotor are assigned an isotropic hysteresis material (M330-35A). The stator coils are excited with 3-phase sinusoidal currents and a static simulation was performed for 90 rotor positions. 
Two simulation cases are considered: case (i) evaluates hysteresis and excess losses without classical eddy-current coupling, and case (ii) additionally includes eddy-current losses at the lamination scale. The generalized model is evaluated using two approaches. The online approach computes potential, gradient, and Hessian values during simulation whereas the offline approach retrieves precomputed values via interpolation, resulting in a negligible maximum interpolation error ($\approx$$0.02\%$). The simplified model is evaluated with the online approach only.
In case~(i), the simplified model achieves $88\%$ and $61\%$ runtime reductions compared to the generalized model using online and offline approaches, respectively. The maximum relative difference (relative to generalized model with offline approach) in hysteresis loss is $\approx$$6.8\%$ (Fig.~\ref{fig_hyst_loss_first_case}), 
while the maximum excess loss difference remains negligible at $\approx$$0.12\%$, since the same excess loss parameters are used for both models. Furthermore, the number of Newton iterations required to solve the nonlinear system of equations was higher at a few rotor positions, reaching a maximum of $8$ compared to $4$ for the generalized model. However, each iteration is computational cheaper due to the simplified material response evaluation, and convergence is achieved with a consistent decrease in the residual.
In case (ii), the maximum hysteresis loss difference reduces to approximately $0.3\%$ (Fig.~\ref{fig_hyst_excess_and_eddy}), as eddy current coupling dominates the field solution and drives both models toward a similar field distribution, resulting in close agreement in the loss values. The maximum relative difference for excess and eddy current losses is $\approx$$0.16\%$ and $\approx$$0.18\%$, respectively.
\section{Conclusion and Outlook}
\label{section4}
In this work,  a thermodynamically consistent simplification of the generalized Prandtl--Ishlinskii stop operator is introduced by retaining the nonlinear anhysteretic response and simplifying the hysteretic part. For isotropic case, this enables a fully analytical return-point mapping, replacing iterative Newton updates with a closed-form solution. Applied to a 2D PMSM FE simulation, significant runtime reductions of~$88\%$ and $61\%$ are achieved compared to the generalized model with online and offline approaches, respectively, with comparable loss accuracy. In future work, we will extend the simplification to anisotropic cases and investigate similar analytical updates.

\ifCLASSOPTIONcaptionsoff
  \newpage
\fi

\end{document}